\documentclass[12pt,preprint]{emulateapj}

\shorttitle{The Solar-System-Scale Disk Around AB Aurigae}
\shortauthors{Oppenheimer et al.}
\journalinfo{Accepted for Publication in The Astrophysical Journal, February 22, 2008.}
\slugcomment{Received October 8, 2007; Accepted February 22, 2008}

\begin{document}

\title{The Solar-System-Scale Disk Around AB Aurigae}

\author{Ben R. Oppenheimer\altaffilmark{1,7}}
\author{Douglas Brenner\altaffilmark{1}}
\author{Sasha Hinkley\altaffilmark{2}}
\author{Neil Zimmerman\altaffilmark{2}}
\author{Anand Sivaramakrishnan\altaffilmark{1}}
\author{Remi Soummer\altaffilmark{1}}
\author{Jeffrey Kuhn\altaffilmark{3}}
\author{James R. Graham\altaffilmark{4}}
\author{Marshall Perrin\altaffilmark{4}}
\author{James P. Lloyd\altaffilmark{5}}
\author{Lewis C. Roberts, Jr.\altaffilmark{6}}
\author{David M. Harrington\altaffilmark{3}}

\altaffiltext{1}{Department of Astrophysics, American Museum of Natural History, 79$^{\rm th}$ Street at Central Park West, New York, NY 10024, USA}
\altaffiltext{2}{Department of Astronomy, Columbia University, 550 West 120$^{\rm th}$ Street, New York, NY 10027, USA}
\altaffiltext{3}{Institute for Astronomy, University of Hawaii, 2680 Woodlawn Drive, Honolulu, HI 96822, USA}
\altaffiltext{4}{Department of Astronomy, University of California at Berkeley, 601 Campbell Hall, Berkeley, CA 94720, USA}
\altaffiltext{5}{Department of Astronomy, Cornell University, 610 Space Sciences Building, Ithaca, NY 14853, USA}
\altaffiltext{6}{The Boeing Company, 535 Lipoa Parkway, Suite 200, Kihei, HI 96753, USA}
\altaffiltext{7}{$^\ast$To whom correspondence should be addressed; E-mail:  bro@amnh.org}

\begin{abstract}
The young star AB Aurigae is surrounded by a complex combination of gas-rich and dust dominated structures.  The inner disk which has not been studied previously at sufficient resolution and imaging dynamic range seems to contain very little gas inside a radius of least 130 astronomical 
units (AU) from the star.  Using adaptive-optics coronagraphy and polarimetry we have imaged the dust in an annulus between 
43 and 302 AU from the star, a region never seen before.  An azimuthal gap in an annulus of dust at a radius of 102 AU, along
with a clearing at closer radii inside this annulus, suggests the formation of at least one small body at an orbital 
distance of about 100 AU.  This structure seems consistent with crude models of mean motion resonances, or accumulation of 
material at two of the Lagrange points relative to the putative object and the star.   We also report a low significance 
detection of a point source in this outer annulus of dust.  This source may be an overdensity in the disk due to dust accreting 
onto an unseen companion.  An alternate interpretation suggests that the object's mass is between 5 and 37 times the mass 
of Jupiter.  The results have implications for circumstellar disk dynamics and planet formation.
\end{abstract}


\keywords{instrumentation: adaptive optics --- 
methods: data analysis --- 
stars: individual (HD31293) 
techniques: image processing --- 
stars: planetary systems 
}


\section{Introduction}
The star AB Aurigae (hereafter AB Aur; distance $d = 144$ parsecs; visual magnitude, $V = 7.04^{\rm m}$; spectral type A0Ve; 
mass $M = 2.4 \pm 0.2$ M$_{\odot}$; age $1-3$ Myr; also known as HD 31293, HIP 22910, BD+30 741; \cite{vdA97,deWarf}) has
 been intensively studied and is widely perceived as prototypical of young, intermediate-mass stars, the so-called Herbig 
Ae stars.  The star belongs to the Taurus-Auriga star-forming region, which is thought to be between 1 and 3 Myr old, where 
the dispersion in ages is most likely due to observational or calibration errors \citep[e.g.]{2007prpl.conf..117W}.   
Because of its brightness, age and proximity to Earth, this star presents an important opportunity for investigating star 
and perhaps planet formation.  

Indeed, the star exhibits a fascinating array of phenomena, including a reflection nebula extending thousands of astronomical units (AU; the Earth-Sun distance) away from the star \citep{1999ApJ...523L.151G}; a more compact, annular (100-750 AU) 
region mainly composed of gas as detected in $^{13}$CO, $^{12}$CO, C$^{18}$O, HCO$^{+}$, H$_2$, and millimeter wavelength 
continuum imaging \citep[e.g.]{1997ApJ...490..792M,2001Natur.409...60T,1998A&A...336..565H,2005A&A...443..945P}; and an apparent inner region within about 100 AU of the star which exhibits gas depletion, where the ratio of the mass in dust to that in gas is large (based on the surface densities derived by \citep{2005A&A...443..945P}).   It seems that the dusty inner disk has an outer radius at about 130 AU  \citep{2004ApJ...605L..53F}, outside of which the gas column density rises.  Also, near infrared (NIR) imaging reveals a spiral structure 
outside of the inner disk (from 130 to a few hundred AU; \citet{2004ApJ...605L..53F}).  Ancillary to this study, but of importance in understanding this system, the material also exhibits a central hole of radius between 0.3 and 0.6 AU \citep{2001ApJ...546..358M}.  A bright spot has also been detected about
 1-4 AU away from the star \citep{2006ApJ...645L..77M} with NIR interferometry.  
 
Interpretations of these observations are 
inconsistent with each other in the literature, with some authors describing disk-like structures and others not.  The system is clearly complicated, and it may have properties of both gas-rich and debris disks, based on qualitative consideration of the observations mentioned above.  As such, it may be in a stage of evolution that is intermediate between these two states, and it presents an ideal observational opportunity to study disk evolution and the formation of small bodies in such disks.  This means 
that high resolution observations of the region between 1 and 130 AU are particularly important.  

\begin{figure*}[ht]
\center{\includegraphics[height=6in]{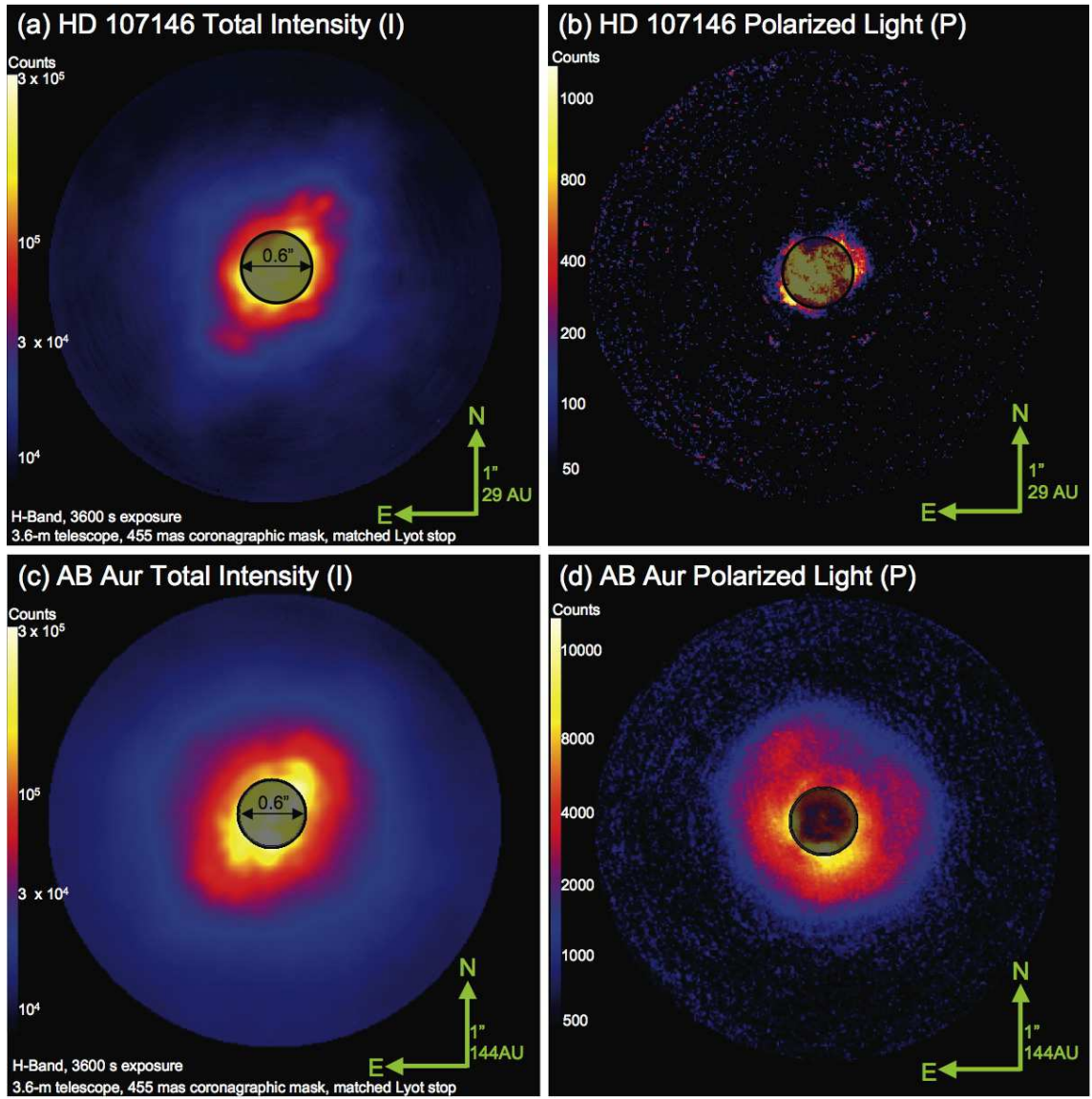}}
\caption{Images of the stars HD 107146 (a and b) and AB Aur (c and d) showing the 4.7 arcsecond field of view with N up and E left.  (a) and (c) are the summed intensity I images shown with logarithmic scale spanning the minimum and maximum counts.  Significant starlight exists throughout the field of view for both stars.  (b) and (d) are the double-differenced $P$ images (see text) also shown with logarithmic color scale.  The coronagraphic occulting spot is slightly smaller (0.45 arcsecond diameter) than the rejected region of each image, indicated by the shaded circle 0.6 arcsecond in diameter in both images. Cardinal direction arrows are 1 arcsecond long and drawn to scale. Note that the maximum count level in (d) is about ten times that in (b).  }
\end{figure*}

However, this region has been difficult to image in light scattered from the dust because of the extreme brightness 
difference between the star and the dust.  We present the first images of polarized light from this region in the NIR with 
sufficient contrast (residual starlight after suppression of $\le 10^{-5}$ at radii of 0.3 to 2.2 arcseconds) and resolution 
(14 AU) to resolve structure in the dust disk.  This level of starlight suppression was achieved using polarimetry in conjunction 
with high-order adaptive optics (AO; for spatial resolution) and advanced coronagraphy (for starlight suppression) under 
the aegis of the Lyot Project \citep{2004SPIE.5490..433O,2003fst3.book..155O,2007ApJ...654..633H,2001ApJ...552..397S}.  

\begin{figure*}[ht]
\center{\includegraphics[width=6in]{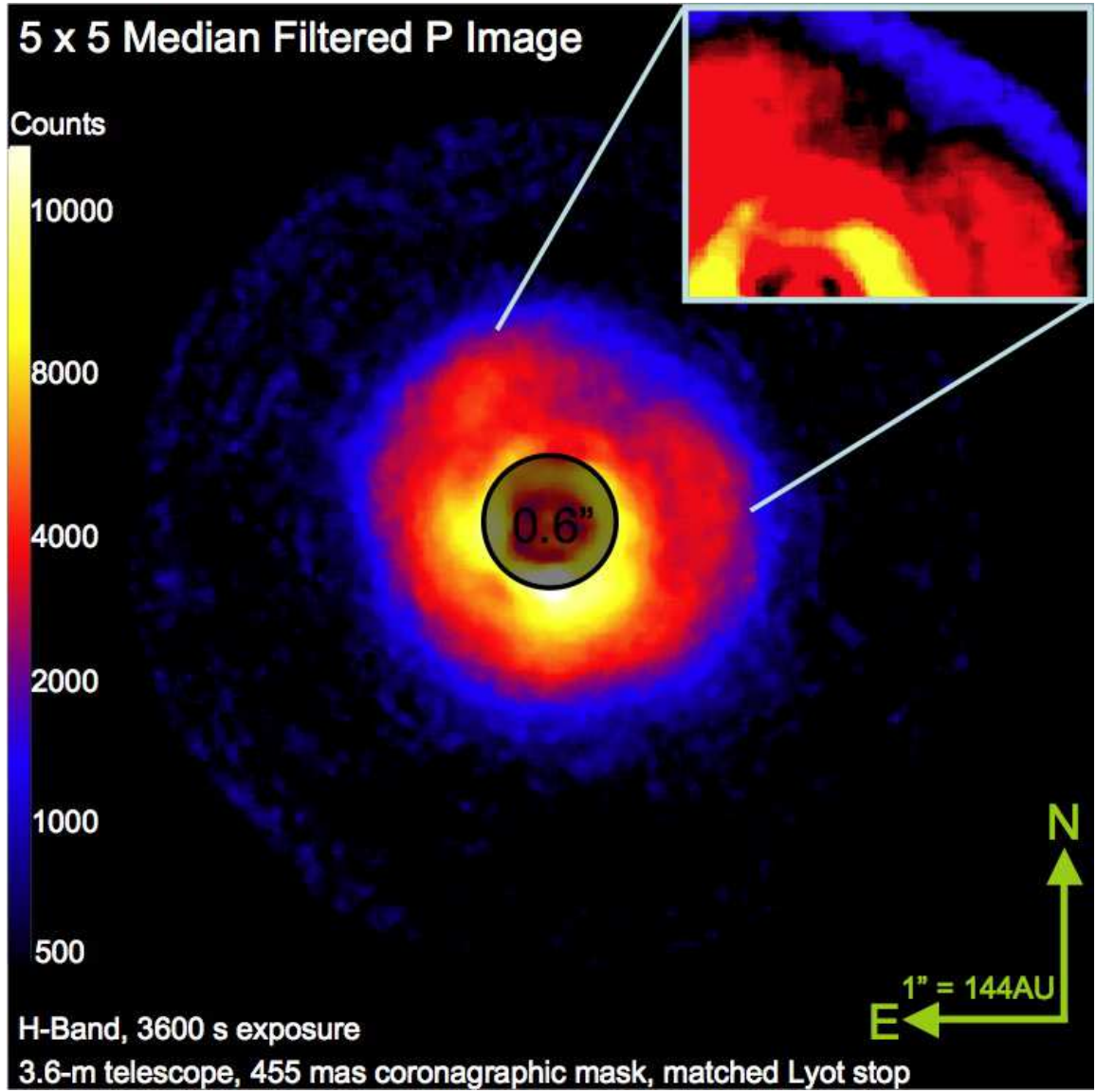}}
\caption{$P$ image of the AB Aur system as in Fig.~1d, but with a median filter applied with a five pixel bounding box, matched to half the diffraction-limited resolution of the instrument.  
The image at upper right is a magnified portion centered on the location of a point source (see text).  In this box, the color scale is adjusted to aid interpretation.  The color scale is logarithmic.} 
\end{figure*}

\section{Polarimetry}
Scattered light from the gas and dust in the vicinity of a star should be linearly polarized with 
the electric field perpendicular to the plane of incidence as seen from the telescope (assuming a single scattering 
process). Thus, dust in close proximity to a star will, in general, induce a spatially variable polarization angle and fractional polarization 
as seen in a polarized light image.  This makes imaging polarimetry of stars a valuable tool for studying star formation.  
In fact, numerous imaging polarimetry observations of disks have been made, permitting the observation of extremely faint sources
 in close proximity to stars, because starlight can be efficiently removed by forming a polarized light image 
\citep{2001ApJ...553L.189K,2004Sci...303.1345P,2004A&A...415..671A,2006ApJ...641.1172T}.  Indeed, direct light from the star 
is only minimally polarized by small-angle ($\le 5\times 10^{-4}$ degrees) forward scattering due to the Earth's atmosphere or the 
interstellar medium.  In contrast, light from the star that has scattered off of material in the near-star environment is 
imparted with far more polarization due to the large angle of scattering (near 90$^\circ$ for a face-on disk).  Thus by 
differencing images obtained in orthogonal polarization states ({\it i.e.} $I+Q$ and $I-Q$, $I+U$ and $I-U$, or $I+V$ and $I-V$, using the 
standard Stokes vector notation) the star light is removed, and an image of polarized light, a ``$P$ image,'' can be formed where
 $P=\sqrt{Q^2+U^2+V^2}$. The spatial variation of the P image is a sensitive measure of variations in the scattering 
properties of the near-stellar dust and gas and is insensitive to atmospheric speckle noise. A further discussion can be found in the appendix of this work. The series of images acquired (six sets of two simultaneous images, each with different polarimeter settings; see Appendix) 
are also added together to form a separate total intensity or ``$I$'' image.

\section{Observations and Data Reduction}
Over a period of one hour and thirty-eight minutes, commencing at 09:55 on 2006 December 12 UTC, we acquired ten polarimetric measurements in H-band (1.45-1.83$\mu$m) with the Lyot Project coronograph installed on the 3.63m AEOS telescope atop Haleakala, Maui.  Each measurement is a series of six dual-beam polarized images (60-s exposures) with different liquid-crystal retarder settings (see appendix for details), to permit derivation of the P image.  
The weather at the observatory was photometric, with winds of less than 8 km/h and relative humidity below 12\%.  The local 
Fried parameter, $r_0$, a measure of the strength of turbulence in the atmosphere above the observatory, spanned the range of
 15 to 32 cm, indicating ideal conditions for AO observations at AEOS.  AEOS has a 941-actuator AO system which suppresses 
imaging distortions due to the atmosphere \citep{2002PASP..114.1260R}.  The coronagraph 
\citep{2004SPIE.5490..433O,2003fst3.book..155O,2007ApJ...654..633H} is coupled with a NIR camera called 
``The Kermit'' \citep{PerrinAMOS03}, with which we image the 4.7 arcsecond field of view centered on the star at a pixel scale 
of 13.7 mas pixel$^{-1}$.  The diffraction-limited imaging resolution of the system is approximately 94 mas in $H$-band.

Used here as an experimental ``control'' observation, data on HD 107146 (a $V=7.07^m$ G2V star) were taken in a manner identical to that of AB Aur commencing at 15:14 on 2006 December 12 UTC, with a similar range of weather conditions and telescope pointing relative to the horizon. We observed this star because it has a known debris disk, although the likelihood of our detecting it was extremely low, because the disk is optically very thin and has only been imaged outside of the Lyot Project field of view \citep{2004ApJ...617L.147A}.  Furthermore, a lack of infrared excess at 18 $\mu$m indicates a central hole at least as wide as 31 AU (1 arcsec) in radius \citep{2004ApJ...604..414W}.  As expected, HD 107146's disk was not detected within our field of view (within 67 AU). Further discussion of these points can be found in the appendix. 
 
Calibration of the system includes standard dark current and flat-fielding gain images (acquired using incandescent lamps each night), bad pixel maps \citep{2006EAS....22..199S,2007ApJ...654..633H}, and observations of binary stars with well-known orbits to ascertain pixel scale and image rotation fiducials. The appendix to this work describes the steps in more detail.   

Fig.~1 shows the $I$ and $P$ images of both HD 107146 and AB Aur.  The structure we see in these images is better analyzed after filtering the image with a median pixel selection over a box matched to half the instrument's diffraction limit (3.5 pixels, rounded up to 5).  Oversampling in the image plane aids speckle removal in postprocessing, but it can dilute real signals in final analyses.  The median filtered $P$ image of AB Aur is shown in Fig.~2.   Before analyzing the structure in this image, we must establish quantitative sensitivity limits.

\section{Detection Limits}
According to a technique we developed previously \citep{2007ApJ...654..633H}, we have derived the difference in magnitude between the occulted star and a 3-$\sigma$ point source as a function of position in the $I$ images for HD 107146, to evaluate the sensitivity of the AB Aur data without the presence of its disk, or any other source in the field of view.   

\begin{figure}[ht]
\center{\includegraphics[angle=0,width=3.3in]{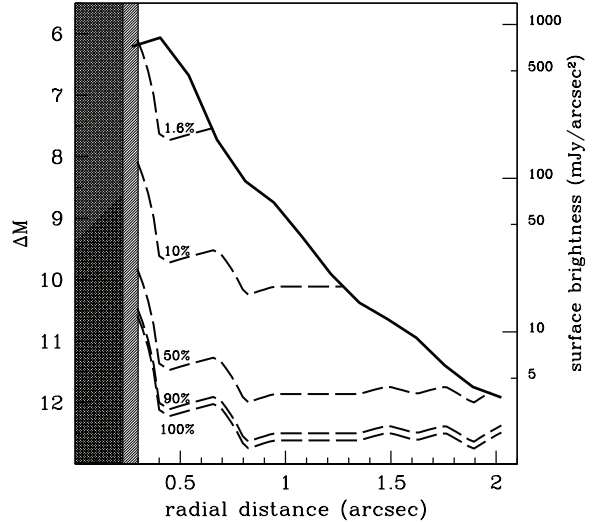}}
\caption{Dynamic Range Plot.  This plot shows the difference in magnitude ($\Delta M$) between a star occulted by the coronagraph and a 3-$\sigma$ point source or surface brightness as a function of angular separation.  Corresponding surface brightness is given on the right.  The size of the occulting mask is indicated by the dark shaded region and the lighter shaded region corresponds to the rejected part of the data.  The solid line indicates the dynamic range for the $I$-image while the dashed lines are dynamic range functions indicating the {\it total} intensity for sources with various fractional (\%) polarizations.} 
\end{figure}

Fig.~3, a so-called ``dynamic range plot'' \citep{2000SPIE.4007..899O}, shows the difference in magnitude ($\Delta M$) between the occulted star and a point source that would be detected at the 3-$\sigma$ level in our $I$ image, as a function of radial distance from the star.  The photometric values were calibrated to unocculted images of AB Aur taken immediately prior to the occulted images.  An extension of this technique was applied to the polarimetric data to derive the sensitivity limits for various fractional polarizations.  For a real source in our data, only a lower limit to the fractional polarization is measurable, because of the contamination from the starlight throughout the image (see appendix).

\section{AB Aur Images}
The images of the AB Aur system (Figs.~1c, 1d and 2)  exhibit several important features.  In the $I$ image, starlight dominates the entire field of view (Fig.~1c) and shows some elongation along a position angle (PA) of 321 $\pm$ 1$^\circ$.  This elongation, which is purely instrumental, and apparent in both the AB Aur data and the HD 107146 I -image, is due to a combination of the wind's effect on AO image improvement \citep[e.g.]{2006ApJ...647.1517S}, and some residual image motion compensation errors.  (The wind had an average prevailing direction of 320$^\circ$ during these observations.)  In the $P$ image (Fig.~1d), light is detected throughout the field of view but is mostly constrained to a region 2.1 arcseconds (302 AU) in diameter centered near the star.  Outside of the central region, the detection is of low significance, but does exhibit some amplification to the NE, which is where the brightest ``spiral'' feature has been seen in other studies \citep{2004ApJ...605L..53F}.  In the inner region significant structure is present.   The photocenter of the inner disk is shifted 88 mas (12 AU) along a PA of 336 $\pm$ 2$^\circ$.  The outer annulus of material in this inner disk shows a distinctly non-uniform distribution in the azimuthal direction, with a deep depletion of polarization centered 700 mas (102 AU) away from the star at a PA of 333 $\pm$ 2$^\circ$.  In a study conducted at the Subaru telescope \citep{2004ApJ...605L..53F}, there appears to be a similar depleted region in this outer annulus, although it is right on the edge of their coronagraphic mask.

It is important to note that none of the structure seen in the $P$ image correlates with structure in the $I$ image, reinforcing the fact that only lower limits to the fractional $P$ can be determined. For example, the $I$-image elongation is at least 15 degrees away from the depleted region's azimuthal angle in the outer annulus.  The residual $P/I$ amplitude in HD 107146 was about 1\%. In contrast the $P/I$ structure we observe in AB Aur was between 1 and 9\% --- significantly larger than the residual systematic polarization noise we measure in the non-detection around HD 107146.  

In the center of the depleted region, there appears to be a 2.8-$\sigma$ amplification of polarization fraction spanning a seven-pixel diameter circular aperture ( $\sim \lambda / D_{\rm Lyot}$, or 94 mas).  This appears to be an unresolved point source with $P/I\ge 5$\%.  We investigated all other features in the image and found none to have point source shape or scale (based on unocculted images) with significance greater than 2.5-$\sigma$.  No optical ghosts have ever been observed in the NIR in our system since it was commissioned in early 2004, and we have never observed spurious reflections from bright stars outside the field of view.  This putative point source may or may not be real, and despite its marginal detection, we include it in our analysis below, although the principal observations of structure in the disk stand regardless of the reality of the point source.

\begin{figure}[ht]
\center{\includegraphics[angle=0,width=3.3in]{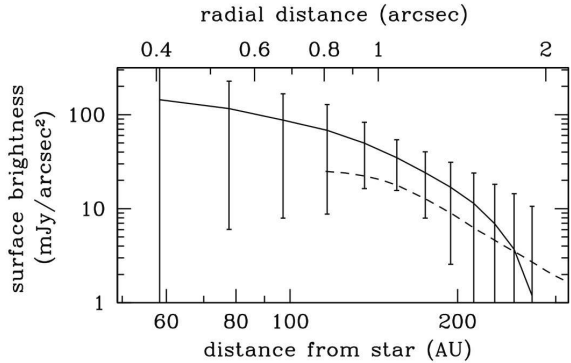}}
\caption{Derived radial surface brightness of the AB Aur disk in Stokes $I$, using the HD 107146 data, where no polarized diffuse light was detected, as a reference azimuthally averaged brightness of an occulted point source.  Error bars indicate one standard deviation for the range of values observed in each annulus of data.  The dashed line indicates the radial profile derived from Subaru coronagraphic data \citep{2004ApJ...605L..53F}.} 
\end{figure}

Further structure in the P image seems to indicate a drop in polarized light as one proceeds radially inward from the outer region toward the star.  This is particularly pronounced in the NW at a radius of about 570 mas (83 AU).  Closer to the star, the polarized light increases again up to the region obscured by the coronagraphic mask.  There may be an annular depletion zone in close proximity to the coronagraphic mask edge, but we state this with less certainty.  This annular depletion zone (which, also, shows no correlation with speckles in the $I$ image) is located at roughly 300 mas (43 AU) radius from the star and a PA of about 142$^\circ$.  

\section{Interpretation: Circumstellar Material}  The azimuthally averaged surface brightness of the circumstellar material provides initial measurements of the disk material.  We use an azimuthal average, because the only effective way to remove the stellar light from the I-image, without restricting study to polarized light, is to use a reference star with no detectable circumstellar material.  AO point spread functions are highly variable and complex, meaning that point-spread function subtraction is not practical (at best).  An azimuthal average permits a first order understanding of the difference between AB Aur and a reference star, and the simplest measure of flux from the disk alone.  To this end, we used HD 107146's azimuthally averaged $I$-image subtracted from the same average for AB Aur (Fig.~4).  The disk shows surface brightness ranging from about 80 to 3 mJy/arcsec$^2$ at radii of 50 to 150 AU.  This agrees with the values previously derived from 120 AU outward in the H-band \citep{2004ApJ...605L..53F}.  These values are also useable as a conservative upper limit to HD 107146's disk flux in the 8 to 66 AU region around that star.

Interpretation of the morphology present in the P image depends on the notion that that polarization fraction in a given pixel in the image is a function of the column density of dust in the disk.   What exactly this means in terms of the disk's structure is not entirely clear.  A number of other studies suggest that this disk is optically thick in the mid-IR out to radii of about 118 AU \cite{2006ApJ...653.1353M}, roughly the region we have detected, although data on the spatial scales reported here are lacking.  If the disk is optically thick at the wavelengths we are observing here, then the P image reveals only a map of the inhomogeneities in the surface of the disk, not the mid-plane features.   If the disk is optically thin at these wavelengths and spatial scales, then the P image will map the full structure of the dust in the disk.  This is critical to interpretation, because, as \cite{2007ApJ...659L.169J} note, the surface structure in a relatively thick disk will not necessarily reflect the interior density structure.  The full structure of the density is necessary to utilize detailed modeling of disks to provide rigorous theoretical interpretation.  therefore, we need to know whether the disk is optically thick or thin.  However, the estimates of optical depth derived thus far are based on extremely simplistic models that assume azimuthal and vertical homogeneity in the disk structure, with a radial dependence.  What is beyond doubt is that these models are inaccurate based on the highly asymmetric morphology of the disk as revealed here.  The conclusion must be that some parts of the disk may be optically thick while others may not be, including some of the regions imaged here.  Numerical models, not simple analytical ones are needed. 

The primary feature of the observed region is in the annulus at an approximate radius of 102 AU with a width equivalent to the image resolution $\pm$ 10 AU, where there is an apparent hole at a PA of 333$^\circ$.  In addition, there are clumps approximately 60$^\circ$ from the gap in the azimuthal direction, and a general reduction of density on the opposite side of the gap.  We note that the very outer edge of this depletion zone is seen in data reporting a spiral structure in the near-IR, right at the edge of the corresponding coronagraphic mask \citep{2004ApJ...605L..53F}.  This structure, as noted above, also seems offset from the stellar position by about 12 AU.  Since we are imaging in polarized light, some of this geometry could be due to the efficiency of scattering due to the inclination angle of the disk, variously reported for this inner region to be between 12$^\circ$ and 30$^\circ$, with a major axis near 60 to 80$^\circ$ \citep{2006ApJ...653.1353M}.  Indeed, if these numbers are correct, the annular depletion zone is nearly aligned with the minor axis of the disk and the opposite side would be expected to have reduced polarization fraction because it would be subject to the more weakly polarizing forward scattering by the dust.  Because of the great uncertainty in the inclination angle and the large range in values depending on the material imaged in a given study, we did not see the value in adopting a particular inclination angle for the region we have imaged.  Thus we have no ``deprojected'' image of the disk, which would also require a model for the scattering material and its Mueller matrix. 

A veritable zoo of models for structure in circumstellar disks exists, with a dominant competition currently between so-called ``core accretion'' and ``gravitational instability'' theories to explain the formation of low mass objects in disks.  These competing models predict different types of structure in disks.  However, as far as could be determined from published material, the structure we see in this disk is most similar to models where a small object is present that dynamically induces mean motion resonance structure \citep[e.g]{2007P&SS...55..569W}.  That structure creates amplifications of density in the azimuthal direction roughly $\pm60^\circ$ away from the forming object, with a weaker clump opposite the object \citep[{\rm esp. Fig.~10}]{2007P&SS...55..569W}.  This structure also has clearing just inside the annulus as well as amplified density further in, another set of features that seem to exist in our $P$ images.  
There are issues with interpreting AB Aur's disk in this manner.  These models exhibiting mean motion resonance generally require optically thin disks with no gas present.  We made the point above that the optical depth is uncertain, and other observations show that this inner region is highly depleted of gas \citep{2005A&A...443..945P}.  If this interpretation is correct, a small body would be forming in the gap and should be centered therein.  However, the theoretical modeling effort for disks of this type is not sufficiently mature to provide a complete interpretation.  The observations can most simply be interpreted as showing an amplification of material around the L4 and L5 Trojan or Lagrange points relative to an object that is situated in the gap.  We discuss the possible detection of that object below.

Comparing the AB Aur structure with other disks is important.   First we note the large size of this disk and associated other structures ($\ge 120$ AU). A few low-mass companion formation theories exclude the formation of objects at such large distances as 102 AU, due to a presumption of very low dust density \citep[e.g. {\rm and references therein}]{2007ApJ...659L.169J}.  These models claim that very low mass companions at such distances could only be there through dynamical interactions after forming near the star, and therefore should have highly eccentric orbits.  The core accretion models can form objects at such distances, but generally of extremely low mass \citep{2005SSRv..116...67B}, and interestingly, the gravitational instability models can only form objects in the 50 to 100 AU range because the disk is too hot closer to the star \citep{2007ApJ...659L.169J}.  However, observations  of other disks reveal significant structure at scales certainly as large as this, albeit for older systems.  For example, the star Fomalhaut has an offset ring of material, with some models claiming that this ring is due to a Saturn mass object at an orbital distance of $\sim120$ AU \citep{2006MNRAS.372L..14Q} with low eccentricity.  In some respects one could speculate that the AB Aur structure is an evolutionary precursor to that found around the older Fomalhaut, which is very similar in physical properties to AB Aur except for its age.  Furthermore, the star HD 142527 shows evidence for another offset ($\sim20$ AU) disk of radius 500 AU and a hole of about 125 AU in radius \citep{2007lyot.confE..46F}.  Such a structure is much larger than what we report here, and not too dissimilar to a smaller scale structure around epsilon Eridani \citep{1998ApJ...506L.133G}.  

Unfortunately few other Herbig Ae/Be stars have been observed at the spatial scales ($\le$ 100 AU) presented here or with polarimetry, so direct comparison with similarly aged objects is difficult.  There are two possible exceptions: observations of HD 141569 and HD 100546, which have been observed with, for example Hubble STIS on larger scales and in optical scattered light (which traces a variety of phenomena; \cite{2005ApJ...630..958G,2001AJ....122.3396G}).  In the case of HD 100546, complex structure is revealed at essentially all probed radii from the star and because it is significantly fainter at larger radii ($\ge 500$ AU) from the star than AB Aur, it may be more evolved and optically thin.  The observations of this star within 100 AU are, unfortunately lacking and so direct comparison with these observations provides little insight.  

For HD 141569, which is widely described as a ``transitional'' disk \citep[e.g.]{2005AJ....129.2481Q,2007A&A...476..829H}, shows a spiral structure, also at large radii ($\ge$ 300 AU) and a continuous distribution of dust at closer radii with multiple ring-like structures \citep{2003AJ....126..385C}.  This has direct similarity with the structures we have observed in AB Aur, and the star is of similar spectra type and age.  The azimuthal gap structure that we observe, though, does not seem to be observed in any other circumstellar disk material.

One other comparison can be drawn with the star 49 Ceti \citep{2007ApJ...661..368W}, which is an A1V star about 10-20 Myr old.  Some authors claim that this object is evolving from the Herbig Ae stage to the more mature ``Vega'' state, and it is clear that there is a transition in dust properties with smaller grains dominating in the 30-60 AU region \citep{2007ApJ...661..368W}.  This may be a slightly more mature version of the disk we observed here and shows a similar radial hierarchy of structures and dust properties, even though in AB Aur the transition in dust properties seems to be at 130 AU, rather than at 60 AU.  The difference in age suggests that these very similar stars, when compared, provide clues to the evolution of disks around intermediate mass stars.

\section{Interpretation: A Probable Companion}
It may be possible to derive a mass for whatever object is causing the perturbations in the dust density in the disk through detailed modeling of the resonance structure and an estimate of the mass of material in this structure.  This is beyond the scope of this paper, and requires significant theoretical work.  Here, we explore the possibility that the point source reported above in the annular depletion zone is a {\it bona fide} companion of AB Aur, and provide two possible interpretations of what it is.  This discussion assumes that the detection is real, even though it has low signal-to-noise ratio.

Although we have not confirmed that the object detected in the gap is physically associated with AB Aur, through the detection of orbital motion or common proper motion, we have determined the probablilty of fortuitous alignment with unassociated background sources. More details on this calculation are provided in the appendix. The object's apparent $H$ magnitude, $m_H$, ranges from 12.76 in the brightest possibility where it is just below the detection limit in the I-image (at a radius of 0.7 arcsec), and 17.26 in the faintest case where it is 100\% polarized.  These are the sum of the values from the dynamic range curve and $m_H = 5.06^m$ for AB Aur \citep{2006AJ....131.1163S}.  Our model indicates that the probability that a background object happens to lie in this location is less than $3.4\times10^{-7}$ for the brighter limit and $9.9\times10^{-5}$ for the fainter one.  The probability is so minute that we assume physical companionship.

There are two ways to interpret this companion.  First, we can derive limits to its mass by assuming that we have detected thermal emission from a cooling body coeval with the star.  The $H$-band ($\lambda = 1.6 \mu$m) absolute magnitude based on the distance to AB Aur is $7.0^m \le M_H \le 11.5^m$.  Using a solar luminosity of $M_H = 3.35^m$, the H band luminosity of the object is between 3.5$\times 10^{-2}$ and 5.6$\times 10^{-4}$ L$_{\odot}$, the luminosity of the Sun.  Given the 1 to 3 Myr age of the system, this companion has a mass of between 5 and 37 M$_{\rm J}$ based on theoretical cooling models \cite{2005SSRv..116...67B}.  Although these models are not ideal for such young ages and the results are dependent on the initial conditions for the models, there is a lack of alternative theory to constrain these physical parameters.

A second interpretation may be more appropriate in this case, because the object exhibits polarization of greater than 5\%.  Although planets orbiting stars should exhibit polarization, when viewed in reflected light, as high as 50\% \cite[e.g.]{1978Icar...33..558T} there is no possibility, due to the brightness, that this source represents light from the star reflected off of a tiny planet at 100 AU.  In fact, if it were reflected light, the simple geometry of the reflection would demand that the object have a radius of 0.1 AU, about 10 times larger than the primary star, if the albedo were 1.  The object is in the midst of a rather dusty region, albeit depleted from neighboring regions.  Perhaps our images merely detect the infall of dust onto an obscured, accreting companion, rather than the companion itself.  Models of planet formation suggest, in some cases, vortices of material surrounding the sites of cores of new planets.  Such regions would be unresolved at the imaging resolution available to the Lyot Project, for example.  As such, they would appear to be unresolved clumps of dust that could scatter starlight just as the main part of the disk would.  Other models, where the disk is denser (optically thick even), predict bright spots and shadows at or slightly offset from the location of planet accretion due to a hole formed by the influx of material onto the forming body \cite[e.g.]{2007lyot.confE..18J}.  No models exist that permit us to estimate a mass for this object based on the polarimetry presented here.  However, if this is an accretion process, one would certainly expect a high degree of temporal variability as the companion is bombarded at random moments by larger particles or planetessimals as the disk continues to thin out.  

\section{Conclusion}
AB Aur and its environment are indisputably a complex system that provide clues to the physical processes of star and planet formation.  Our study provides some insight into a new scale on par with our own solar system.  The combination of observations of this system suggest that (a), because of the 1-4 AU gap between the star and disk \citep{2001ApJ...546..358M,2006ApJ...645L..77M}, the star is mostly formed and has completed its main accretion phase, although there is significant outflow still; (b) planet formation, if it will happen, has already begun, due to the presence of an apparently organized disk with depleted gas \citep{2005A&A...443..945P}, some of which may be adsorbed onto the dust itself; (c) the region imaged here should have a similar temperature to the giant planet region of our solar system, and as such these data could provide constraints on giant planet formation mechanisms; (d) the remaining thin dust and gas at larger radii are either flowing into or out of the disk, as indicated by strong spiral features at larger distances from the star \citep[e.g.]{2004ApJ...605L..53F,1999ApJ...523L.151G,1998A&A...336..565H}.  In addition, projection effects complicate this interpretation particularly if there really is an organized inner disk (such as we have imaged), with external material distributed more isotropically, or at least not in a series of coplanar disks.

Regarding the structure we have resolved, models of disks showing mean motion resonances also suggest strong outflows or infalls of material at coeval epochs with significant material trapped in mean motion resonant orbits \citep{2007P&SS...55..569W}.  These are the only models that seem to come remotely close to mimicking the structure we observe.  We may be witnessing such a process in action for the first time, or some combination of processes not yet modeled, and the poorly detected point source in our data may be related to the structure in the disk.  Higher resolution and more sensitive observations of this fascinating system may be able to constrain planet and star formation models and permit distinguishing between competing theory.  Unfortunately such observations can only be expected once new, large sub-mm observatories, such as ALMA, are in operation.  The critical need here involves spatial resolution superior to 100 mas and dynamic range greater than 12 magnitudes.  A major space mission such as the Terrestrial Planet Finder, would provide a crystal clear view of what is happening in this fascinating system.  In the near term, HST's NICMOS instrument can conduct similar polarimetric observations at lower resolution, and possibly worse dynamic range.

\acknowledgments
The Lyot Project is based upon work supported by the National Science Foundation under Grant Nos. 0334916, 0215793, and 0520822, as well as grant NNG05GJ86G from the National Aeronautics and Space Administration under the Terrestrial Planet Finder Foundation Science Program.  The Lyot Project grateful acknowledges the support of the US Air Force and NSF in creating the special Advanced Technologies and Instrumentation opportunity that provided access to the AEOS telescope.  Eighty percent of the funds for that program were provided by the US Air Force.  This work is based on observations made at the Maui Space Surveillance System, operated by Detachment 15 of the U.S. Air Force Research Laboratory Directed Energy Directorate.
The Lyot Project is also grateful to Hilary and Ethel Lipsitz, the Cordelia Corporation, the Vincent Astor Fund, Judy Vale, FUTDI, and an anonymous donor, who initiated the project.  BRO thanks Lynne Hillenbrand, Gilles Chabrier, Isabelle Baraffe and Michal Simon for extremely useful comments on early drafts, and Misato Fukagawa for being open and very generous with her data on AB Aur taken with the Subaru instrument. This work has been partially supported by
the National Science Foundation Science and Technology Center for Adaptive Optics,
managed by the University of California at Santa Cruz under cooperative
agreement AST-9876783.

\appendix 
\section{Astrometric Calibration and Data Quality}
We used the star HD 76943 for the astrometric calibration, observed at 14:13 on 2006 December 16 UTC under similar observing conditions.  On this date, the orbit determination shows that the secondary component is at 0.546 arcseconds at a position angle of 333.54$^{\circ}$ (N-E)\citep{2000DDA....31.1402H}.  Images of that star are shown in Fig.~A1 to demonstrate the fidelity of the AO correction, roughly estimated via the Strehl ratio, $S = \exp^{-\sigma^2}$, where $\sigma$ is the root mean square of the residual wave front error.  $S$ exceeded 85\% in these images, as determined through comparison with laboratory-based images with the coronagraph \citep{2004SPIE.5490..504R}.  This performance is similar in acquisition images of AB Aur taken immediately before the primary data acquisition.  

\begin{figure}[ht]
\center{\includegraphics[angle=0,width=3.5in]{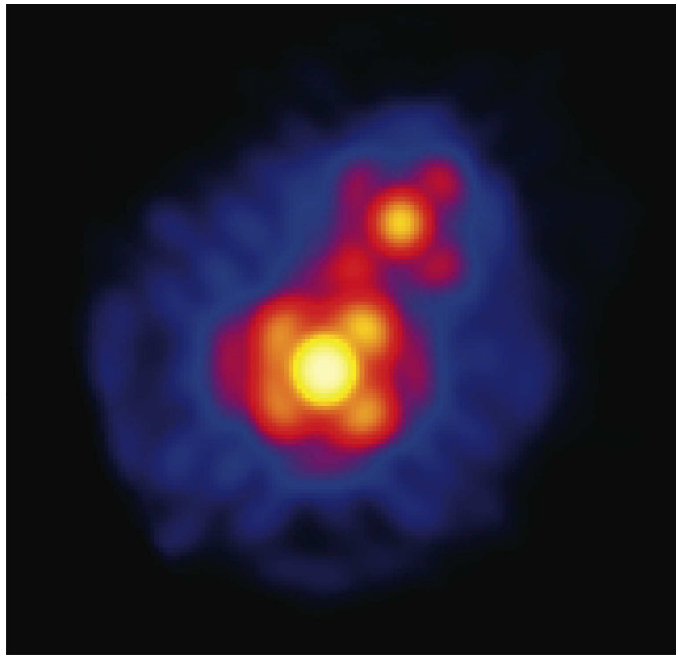}}
\caption{Image of the binary star HD 76943 used to determine plate scale and rotation parameters.  The first Airy ring of the PSF is broken into four spots due to the oversized spider vanes in the coronagraph's Lyot stop.  The secondary star has a separation of 0.546 arcsec at PA 333.54$^{\circ}$.  Pixel scale is 13.7 mas and N is up with E to the left.}
\end{figure}

\section{Polarimetry}
\citet{2001ApJ...553L.189K} describe a double-difference method that minimizes instrumental effects in imaging polarimetry.  To derive our $P$ images, following the \citet{2001ApJ...553L.189K} technique, we take differences of simultaneous orthogonal polarization images ({\it e.g.} $I+Q$ and $I-Q$).  We then difference these images again after spatially interchanging the polarization states \cite{2001ApJ...553L.189K} and then sum them in quadrature to form a ``$P$'' image $P=\sqrt{Q^2+U^2+V^2}.$ Instrumental and other systematic polarization effects are removed or tend to be spatially constant in this polarized image.  To achieve this, a Wollaston prism located immediately after the coronagraph's Lyot stop splits the light into two images, viewed simultaneously side-by-side on the Kermit's detector, with the two images carrying perpendicular vectors of the polarization.  Using liquid crystal variable retarders (calibrated with a Glann-Taylor prism), the instrument can swap the vectors for each field of view or rotate them.  As a result, we obtain six sets of two images.  The two images are recorded on the same detector simultaneously and with the same exposure time.  By swapping the vectors on the two images in a subsequent exposure---the two exposures comprising a ``set''---we can remove instrumental effects induced by the different optical paths of the two simultaneous images.  

Taking the first two exposures, for example, we subtract the left and right images in both and then subtract the resulting final image.  This provides an image in polarized light that we call ``Q,'' with ``U'' and ``V'' images derived with the subsequent exposures in the same way.  These Q, U and V images are from the reference frame of the instrument.  In order to derive the Q, U, V images in the reference frame of the sky, they have to be reprojected taking into account the Mueller matrix of the telescope and instrument as a function of pointing in the sky.  However, this Mueller matrix is not only extremely complex, but is also unknown \citep{2006PASP..118..845H,DavenJeff08}.  Fortunately, the P image is immune to these effects, because it is the quadrature sum of the Q, U and V images, but this "cross-talk" induced by the telescope requires that we also measure V, which is often excluded in observations similar to these.  

Our $P$ image does not directly reveal the absolute polarization amplitude or direction of the near-star optical scattering, because the intensity image has significant spurious scattered starlight and atmospheric speckle noise \cite[e.g.]{2007ApJ...654..633H}.  In addition, as mentioned above, the AEOS telescope changes both the magnitude and polarization angle of incoming light to the telescope \citep{2006PASP..118..845H,DavenJeff08}.  Fortunately this spurious polarization cross-talk is essentially constant over the telescope field-of-view, meaning that the relative structure in the $P$ image is not instrumental in nature.  The observed AB Aurigae Q, U and V images were all very similar in nature with small relative rotations in the polarization angle and degree of polarization as the telescope changes pointing.  However, the quantity $\sqrt{ Q^2 +U^2+V^2 }$ was constant during all ten polarization sequences (6 dual images for each).  Despite these limitations, the spatial variation of the $P$ image is a sensitive measure of variations in the scattering properties of the near-stellar dust and gas and is insensitive to atmospheric speckle noise.

\section{Efficacy of Speckle Suppression} To demonstrate how completely the starlight (and speckle noise) is removed with our imaging polarimetry process, we examine, in more detail, the observations of the star HD 107146  which is used in the main article as a ``control'' non-detection.   It is bright in the coronagraphic I image (Fig.~1a), with a total of 1.33$\times 10^8$ photons detected in a 360 second sequence of exposures.  However, the star is almost entirely absent in the P image (Fig.~1b) with 1.60$\times 10^5$ photons, most of which, due to diffraction, are behind the coronagraphic occulting mask, a region of no consequence to scientific results.  The polarimetry method reduces the residual starlight by a factor of 843 in this case and brings the vast majority of the field of view to the instrumental throughput limits---the $P$ image shows a typical readnoise-limited background.  No speckle structure remains in the region usable for scientific investigations.  Combined with the coronagraph, the central star's light has been suppressed by a factor of about 10$^{-5}$ over the entire field of view, and possibly by much more since our images were not limited by the starlight but rather instrument sensitivity.  It is important to note that none of the speckles in the $I$ image are present in the P image.  This is true for the AB Aur data as well.

\section{Polarimetry Sensitivity} To determine our sensitivity in the $P$ image, artificial point sources with a given intensity and fractional polarization were placed radially into the constituent images prior to composing the $P$ image, with appropriate adjustments in the pairs of simultaneous images so that the relevant fractional polarization was correctly represented by the fake sources. These images were double differenced as described above (Polarimetry).  The total intensity of these point sources in the original images was adjusted until a 3-$\sigma$ detection was obtained in the processed $Q$, $U$, and $V$ data.  ($P$ images have non-Gaussian noise properties so noise must be derived from the constituent images.)  The total intensity (not just the flux in net-polarized light) for a given fractional polarization is plotted in the dynamic range plot, although for any real source detected, only a lower limit to the fractional polarization is measurable, because of the contamination from the starlight throughout the image.

\section{Probability of Point Source Contamination}
We constrained the probability that a point source would be co-aligned with features in AB Aur's disk using our detection limits and a model for the density of objects on the sky.  Deriving a star count-magnitude relationship from the Two-Micron All-Sky Survey Point Source Catalog \citep{2006AJ....131.1163S} and using a fit to galaxy count-magnitude relations \citep{2002ApJ...570...54C}, we find that sources in the region of sky around AB Aur have a sky surface density of
\begin{eqnarray}\log{\frac{dN_s}{dm_H}} = -8.73 + 0.30m_H \\ \log{\frac{dN_g}{dm_H}} = -4.69 + 0.45m_H\end{eqnarray}
\noindent where $N_s$ and $N_g$ are the number of stars and galaxies per square degree, respectively, and $m_H$ is the apparent H-band magnitude.

Because we have detected this source in polarized light, we must modify these equations to reflect the fact that the majority of celestial sources have immeasurable ($\le$ 1\%) polarization with our technique \citep{2004ASPC..309...65W}.  Based on results of a survey of polarized starlight \citep{2002ApJ...564..762F}, we very conservatively assume that 50\% of all stars and galaxies are polarized to the point where they would be detected in our P-image.  This adds a constant of -0.30 to the equations above.

To apply this, we make two estimates of the object's $m_H$, which are 12.76 in the brightest possibility where it is just below the detection limit in the $I$-image (at a radius of 0.7 arcsec), and 17.26 in the faintest case where it is 100\% polarized.  These are the sum of the values from the dynamic range curve and $m_H = 5.06^m$ for AB Aur \citep{2006AJ....131.1163S}.  From equations (1) and (2) and the area of the depleted region in the outer disk annulus, a circle of radius $0.2$ arcsec, the probability that a background object happens to lie in this location is less than $3.4\times10^{-7}$ for the brighter limit and $9.9\times10^{-5}$ for the fainter one.  The probability is so minute that we assume physical companionship.


\clearpage


\end{document}